\documentclass[conference]{IEEEtran}
\IEEEoverridecommandlockouts
\usepackage{cite}
\usepackage{amsmath,amssymb,amsfonts}
\usepackage{algorithmic}
\usepackage{graphicx}
\usepackage{textcomp}
\usepackage{xcolor}
\usepackage{caption}
\usepackage{subcaption}
\usepackage{hyperref}
\usepackage{enumitem}
\def\BibTeX{{\rm B\kern-.05em{\sc i\kern-.025em b}\kern-.08em
    T\kern-.1667em\lower.7ex\hbox{E}\kern-.125emX}}
\begin{document}

\title{Domain-Centered Support for Layout, Tasks, and Specification
for Control Flow Graph Visualization}

\author{
    \IEEEauthorblockN{Sabin Devkota\IEEEauthorrefmark{1},
    Matthew P. LeGendre\IEEEauthorrefmark{2}, 
    Adam Kunen\IEEEauthorrefmark{2},
    Pascal Aschwanden\IEEEauthorrefmark{2}, 
    Katherine E. Isaacs\IEEEauthorrefmark{3}}
    \IEEEauthorblockA{\IEEEauthorrefmark{1}The University of Arizona, Tucson,
    USA, Email: devkotasabin@email.arizona.edu} 
    \IEEEauthorblockA{\IEEEauthorrefmark{2}Lawrence Livermore National
    Laboratory, Livermore, CA, Email: \{legendre1, kunen1, aschwanden1\}@llnl.gov}
    \IEEEauthorblockA{\IEEEauthorrefmark{3}The University of Utah, Salt Lake
    City, USA, Email: kisaacs@sci.utah.edu}
}

\maketitle

\begin{abstract}
Computing professionals in areas like compilers, performance analysis, and
  security often analyze and manipulate control flow graphs (CFGs) in their
  work. CFGs are directed networks that describe possible orderings of
  instructions in the execution of a program. Visualizing a CFG is a common
  activity in developing or debugging computational approaches that use them.
  However, general graph drawing layouts, including the hierarchical ones
  frequently applied to CFGs, do not capture CFG-specific structures or tasks
  and thus the resulting drawing may not match the needs of their audience,
  especially for more complicated programs. While several algorithms offer
  flexibility in specifying the layout, they often require expertise with
  graph drawing layouts and primitives that these potential users do not have.
  To bring domain-specific CFG drawing to this audience, we develop
  \textit{CFGConf}, a library designed to match the abstraction level of CFG
  experts.  \textit{CFGConf} provides a JSON interface that produces drawings
  that can stand-alone or be integrated into multi-view visualization systems.
  We developed \textit{CFGConf} through an interactive design process with
  experts while incorporating lessons learned from previous CFG visualization
  systems, a survey of CFG drawing conventions in computing systems
  conferences, and existing design principles for notations. We evaluate
  \textit{CFGConf} in terms of expressiveness, usability, and notational
  efficiency through a user study and illustrative examples. CFG experts were
  able to use the library to produce the domain-aware layouts and appreciated
  the task-aware nature of the specification.
\end{abstract}


\section{Introduction}
\label{sec:intro}
Control Flow Graphs (CFGs) are widely analyzed in areas of computing such as
program analysis, compilation, optimization and
security~\cite{allen1970cfg,ferrante1987pdg,abadi2005cfi}. Representing
possible orders of execution between instructions~\cite{allen1970cfg}, they
are commonly~\cite{isaacs2019graphterm} visualized as node-link diagrams with
general-purpose libraries such as \texttt{dot}~\cite{graphviz}.  By design,
general libraries do not use domain-specific constructs---in this case loops
and functions---in their layout. Furthermore, though large graphs require
filtering to render efficiently, general libraries focus on drawing only, as
{\em meaningful} filtering is domain-dependent. Thus, the resulting drawing
may not match the mental model of the users or aid them with their tasks.

Adapting general layouts to generate domain- and task-specific CFG drawings
requires both considerable knowledge of the layout algorithm and engineering
effort. Users must operate at the level of nodes and edges, rather than domain
structures. Furthermore, one group's use may not completely transfer to the
needs of another group~\cite{devkota2021ccnav}. Thus, despite having
programming skills, network knowledge, and research experience, computing
experts typically do not allocate time to improving CFG visualizations. If the
general layout is not sufficient, they may either toss it aside or spend hours
following labyrinthine lines~\cite{devkota2018cfgexplorer}.

To empower computing experts to more effectively visualize their CFGs, we
designed \textit{CFGConf}, a CFG drawing library with a high-level, JSON-based
specification language.  \textit{CFGConf} provides a concise interface for
domain-aware drawings that are suitable for common CFG tasks and
integrable into larger visualization systems. Our iterative design process
incorporated a survey of CFG drawing conventions, collaboration with domain
experts, existing task analyses, and formative evaluation using
the cognitive dimensions of notation~\cite{cogdimnotation1989}.

We validate \textit{CFGConf}'s expressiveness and usability through
illustrative examples, integration into a multi-view visualization system, a
summative cognitive dimensions discussion, and a user study. Participants in
the study used \textit{CFGConf} to create domain-aware drawings, finding it
generally easy to setup, amenable to their own files, helpful in showing
domain-specific structure, and useful for domain-specific tasks like
filtering. We find the higher level of abstraction assisted systems experts in
creating domain-aware drawings.

\begin{figure}
\centering
\includegraphics[width=0.95\linewidth]{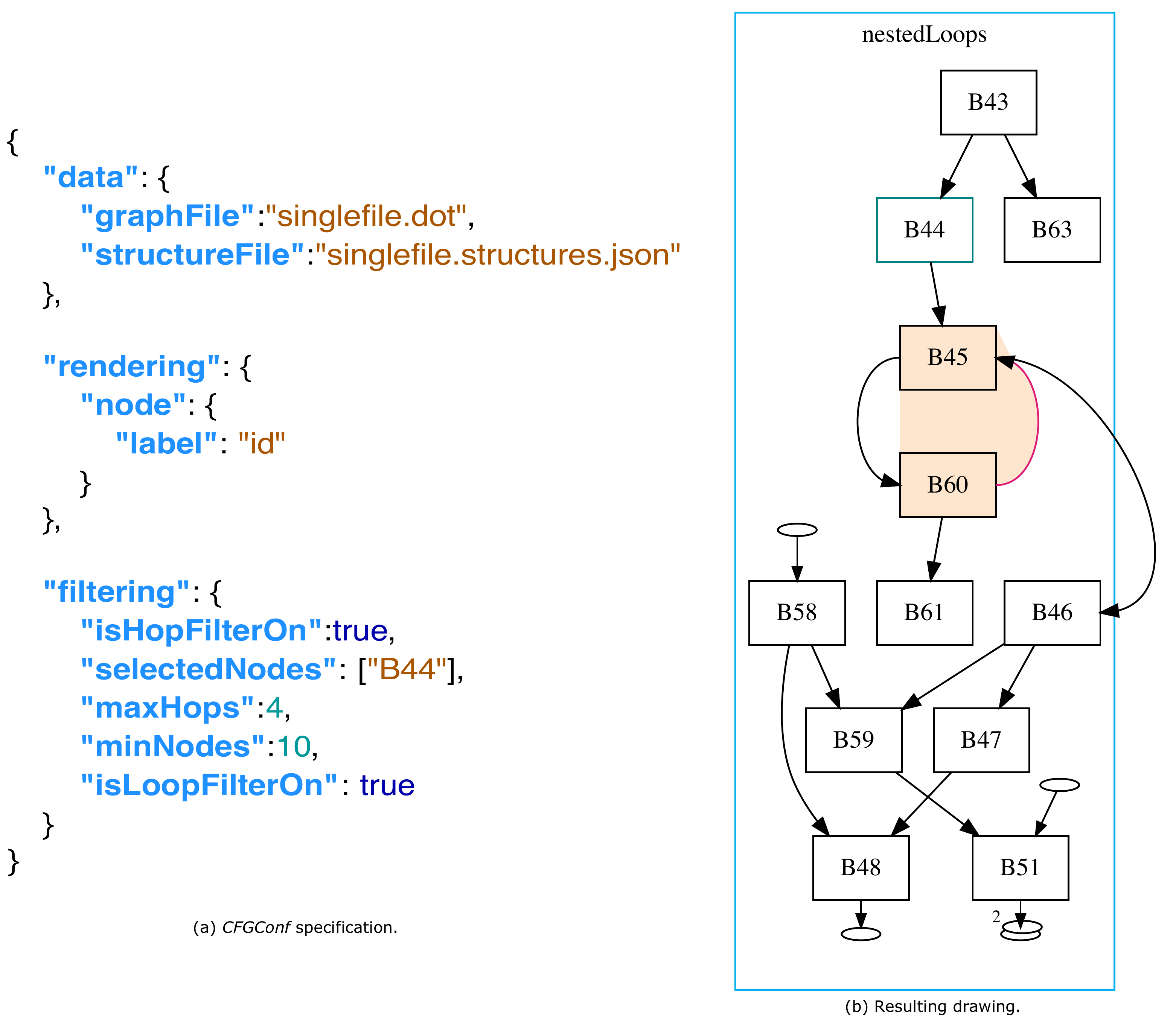}
  \caption{The \textit{CFGConf} specification incorporates domain concepts like
  \textit{functions} and \textit{loops} and domain tasks such as loop-preserving
  filtering to help computer scientists effectively draw control flow graphs.
  Loops are drawn with an orange background color to draw analysts's
  attention. Nodes just outside the filter are drawn with small white discs
  for context.}
\label{fig:teaser}
\end{figure}

\section{Background and Related Work}
\label{sec:background}

A Control Flow Graph (CFG) is a directed graph where the nodes represent
sequences of instructions (i.e., code) that must be executed serially
and edges represent possible paths between those nodes. The nodes are sometimes
referred to as {\em basic blocks}. Edges occur when there are multiple entries
or exits to a basic block, such as due to conditional branching, loops, or
function calls and returns. 

Loops (e.g., \texttt{for}, \texttt{while}) appear as cycles in a CFG. However,
not all cycles are loops. Loops are defined by a {\em back edge} which
connects to the {\em header} node of the loop.

Common analysis tasks when examining a CFG include (a) following the flow of
control (tracing a path through the graph), (b) identifying structures like
loops or functions, (c) exploring details regarding those structures or
regions, (d) understanding those structures in context of the rest of the
CFG, and (e) finding correspondences between the control flow graph and
typical text layouts of code or traces of instructions. See Devkota \&
Isaacs~\cite{devkota2018cfgexplorer} and Devkota et
al.~\cite{devkota2021ccnav} for task analyses of specific CFG visualization
applications.

\subsection{Drawing CFGs}

As CFG tasks often involve path-tracing or identifying graph structures,
node-link diagrams---which support these
tasks~\cite{ghoniem2004,Keller2006}---are the most prevalent idioms. In
particular, Sugiyama-style layouts~\cite{sugiyama1981} are widely used as a
layout
base~\cite{sander1995vcg,ControlF26:online,balmas2004displaying,wurthinger2008pdg,devkota2018cfgexplorer}.
The main steps of Sugiyama-style layouts assume a directed acyclic graph
(DAG).  Libraries implementing these layouts perform an initial pass to break
cycles by reversing edges, allowing the main DAG layout to run. The edges are
then reverted for the final image. As these algorithms are unaware
of CFG structure, the resulting depiction can order cycle nodes differently
than expected by computing experts.

Some CFG visualizations have used program structures.  For example,
Balmas~\cite{balmas2004displaying} simplified the CFGs by collapsing groups
related to functions or loops in a hierarchical fashion.
CFGExplorer~\cite{devkota2018cfgexplorer} introduces a domain-specific layout
for loops that ensures ordering among loop nodes. Loop membership is encoded
by coloring the background of the loop convex hull.
CCNav~\cite{devkota2021ccnav} integrates CFG drawing into a multi-view system,
filtering the CFG heavily based on highlighting in the other views.  Toprak et
al.~\cite{toprak2014lightweight} aggregate basic blocks by converting the
graph into a regular expression and then drawing that expression with a linear
layout.

While these methods improve upon more general layouts when applied to CFGs,
they limit users' ability to adjust the layout. Several are integrated into
multi-view systems designed for specific applications and therefore may be
cumbersome to adapt for other analyses. Time required to solve
Sugiyama-style layouts also limits the size of the graph that can be rendered,
which is a problem as CFGs of even moderate programs can contain tens of thousands
of nodes. 

\subsection{Specifying Graph Drawings}

Typically there is a separation between graph data and drawing.
GraphML~\cite{brandes2013graphml}, GEXF~\cite{gexf:online}, and
\texttt{dot}~\cite{graphviz} let users specify nodes, edges, hierarchical
groups. Visual styling and the rendering of groups depends on the algorithm
applied.

General graph drawing algorithms often have tuning parameters such as the
number of iterations in a force-directed simulation, but these can be
difficult to relate to graph structure.  \textit{Cola}~\cite{colajs} allows
users to specify drawing constraints between pairs of nodes.
SetCola~\cite{hoffswell2018setcola}, (GD)${^2}$~\cite{ahmed2020gd2}, and
Stratisfimal Layouts~\cite{dibartolomeo2021} allow specifying constraints at a higher
level, for example ring sub-graphs in SetCola and groupings in Stratisfimal
Layouts.  However, these algorithms still require familiarity with graph
drawing criteria and considerations at the node and link level.
\textit{CFGConf} trades generality for the ability of domain experts to use
higher-level abstractions in addition to nodes, links, and groups.

\section{CFGConf Design}
\label{sec:design}

The goal of \textit{CFGConf} is to empower people working with CFGs to easily
visualize them in a manner that supports their tasks and mental model. Our
solution is a flexible, domain-specific visual representation and
specification language (Fig.~\ref{fig:teaser}). We designed both through an
iterative process drawing from findings in previous projects, depictions in
computing systems literature, heuristics for notation design, and regular
feedback from domain experts. We summarize our design process and then present
the visual (Sect.~\ref{subsec:visualization_design}) and notational
(Sect.~\ref{subsec:language_design}) design choices. 

\subsection{Design Process}
\label{subsec:design_process}

\textit{CFGConf} was inspired by the following observations stemming from two
different CFG visualization
projects (CFGExplorer~\cite{devkota2018cfgexplorer} and
CcNav~\cite{devkota2021ccnav}),  the domain experts collaborating on them, and
a third party who later used CFGExplorer for their own research.

\begin{enumerate}

  \item Despite both CFGExplorer and CcNav being designed to help
      experts explore compilation and CFGs, task requirements differed enough
      to require different multi-view system designs and CFG drawing
      integration.

  \item The third party used CFGExplorer for the CFG layout alone, ignoring
    the other views and features.

  \item Though the system design needs differed among all three user groups,
    all preferred the loop-semantic layout. This was the reason the
    experts in the CcNav project approached our visualization team.

\end{enumerate}

These observations suggest a need for domain-aware CFG visualization, both in
stand-alone scenarios such as the third party's use of CFGExplorer and for
integration into other visualization systems to support more complicated tasks
and workflows. To support this need, we set out to design a flexible CFG
drawing library. 

We first reviewed the task analyses and drawing needs gathered in the
CFGExplorer and CcNav design studies. For a more general perspective, we then
reviewed example drawings from diagrams in academic papers (described below)
to verify and extend our list of drawing conventions. Next, we proposed an
initial set of drawing conventions and an initial specification to our domain
expert collaborators. We met every other week over a period of eight months to
prioritize and refine aspects of the visual design and specification.

Our domain expert collaborators, who are authors on this paper, include an expert
in program analysis in high performance computing (HPC), an expert in HPC
applications who analyzes program compilation for performance and portability,
and a software developer who hardens HPC tools. These collaborators also
served as domain experts in the CcNav project. However, \textit{CFGConf} is
not an extension of either CFGExplorer or CcNav, but a stand-alone library
intended to ease the creation of such systems that include CFG drawings as
well as enable stand-alone exploration of CFG drawings.

\textbf{Literature Review of CFG Drawing Conventions.} To find drawing
conventions used by a broader set of experts, we examined figures in research
papers published in high profile conferences in compilation, programming
languages, and HPC. We started from the most current proceedings and examined
each paper, working backwards in time until we determined we were not seeing
anything different, suggesting saturation. This resulted in 120 figures from
65 papers. We coded drawing conventions in each graph-like diagram found, with
the exception of tool architecture diagrams. We summarize our findings here.
A detailed discussion and list of papers are in the 
supplemental materials (\url{https://osf.io/vm3pc}).

We found the majority of CFGs and related graphs (e.g., program dependence
graphs, data flow graphs, etc.) followed a top-down ordering of dependencies
with labeled nodes. Half the CFGs had canonical loop layouts as we use here
(but without background coloring or back edge highlighting), even
though some diagrams did not contain loops. There were a variety of edge
drawing styles, with a preference for straight or curved edges over ones with
hard bends. Many of the graphs also had labeled edges or used an enclosing
boundary to indicate function or other grouping behavior. 

The diagrams had a variety of node styling options including color, border,
and shape. Many used edge weights and dashed styles. In \textit{CFGConf},
users can set these styling options globally and override
them for individual nodes or links.

\begin{figure*}[tbp]
\centering
     \begin{subfigure}[t]{0.98\textwidth}
         \centering
         \includegraphics[width=0.98\textwidth]{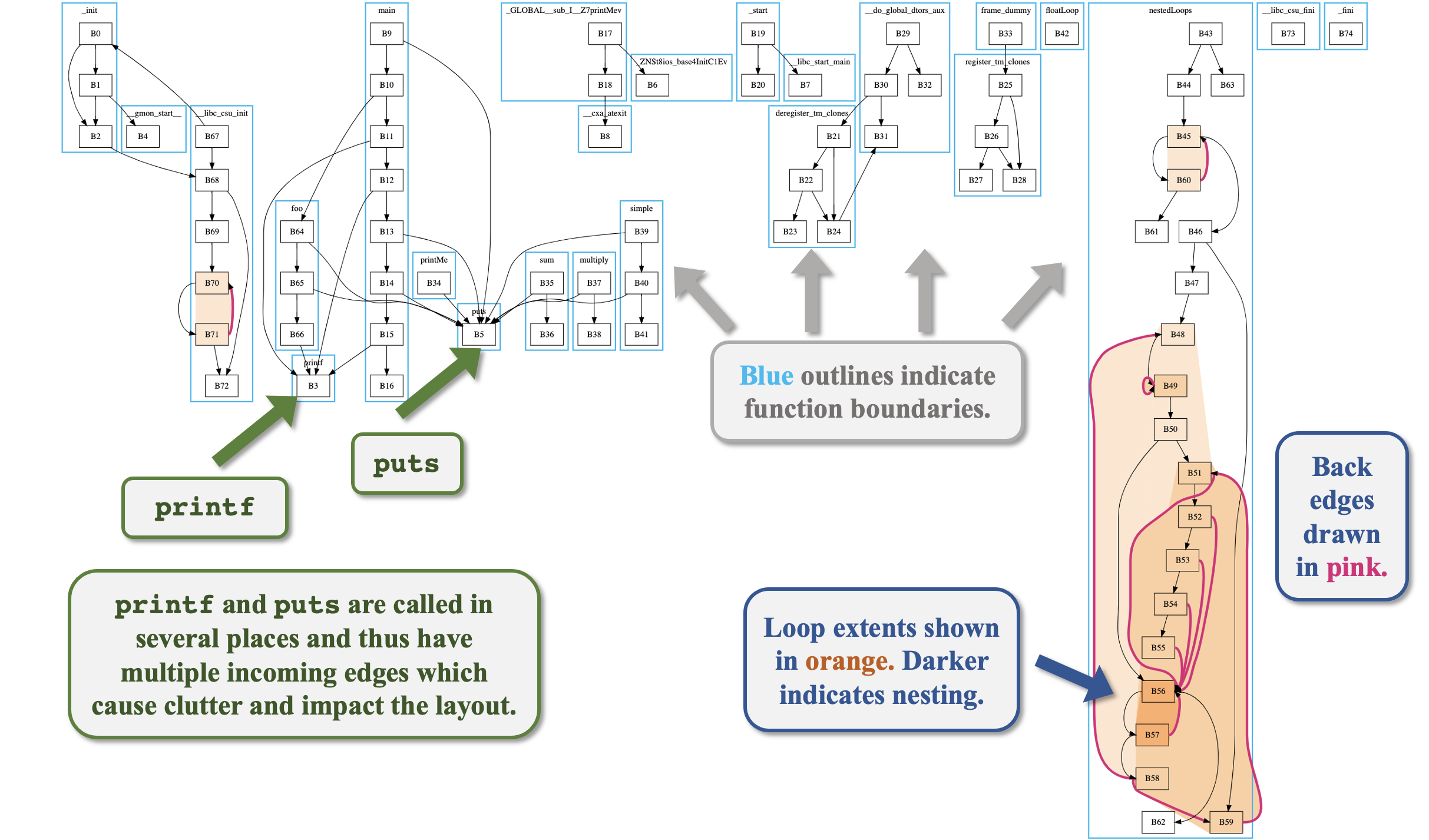}
	 \caption{CFG with function boundaries and no filtering.}
         \label{fig:fxnseverywhere}
     \end{subfigure}
     
     \vspace{3ex}

     \begin{subfigure}[t]{0.98\textwidth}
         \centering
         \includegraphics[width=0.98\textwidth]{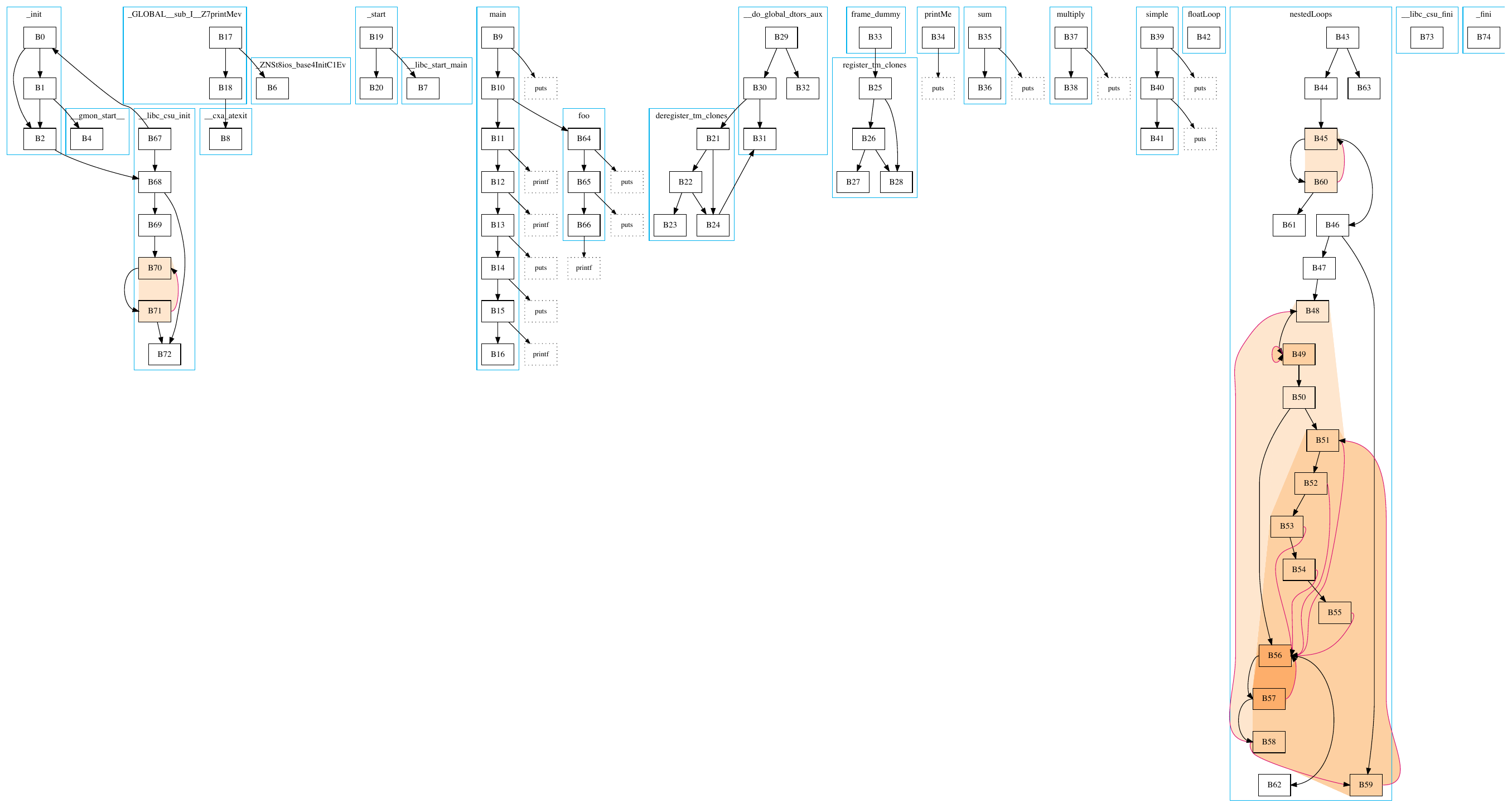}
	 \caption{CFG from above with collapsing of auxiliary functions
	 \texttt{printf} and \texttt{puts}.}
         \label{fig:fxncollapsing}
     \end{subfigure}
\caption{CFG from a small program. We have annotated the top figure to point
  out several features. No filtering has been applied, but the data
  is collected using \texttt{optparser}~\cite{devkota2021ccnav} which elides
  call-return edges. The top shows the whole CFG. The bottom applies the collapsing feature:
  the functions \texttt{printf} and \texttt{puts} are collapsed, duplicated,
  and drawn with dotted boundaries, removing the tangle of lines accessing
  those functions.} 
\label{fig:singlefile}
\end{figure*}

\subsection{Visualization Design}
\label{subsec:visualization_design}

As the literature review and the three groups who used CFGExplorer
or CcNav all support the loop-drawing topology developed in CFGExplorer, we
decided to keep that convention as a central design element in
\textit{CFGConf}. Based on the task analyses and focus on loops, we further
decided to retain the background shading of the convex hull of a loop's nodes 
and the highlighting of the back edge, both exploiting color pop out to make
them more salient.

We then added the following design features, which can be optionally specified
and parameterized in \textit{CFGConf}:

\textbf{Function Boundaries.} Functions are important structures in programs.
We added options to more tightly group nodes of the same function in the
layout and demarcate them with enclosing bounding rectangles, as shown in
Figs.~\ref{fig:teaser} and~\ref{fig:singlefile}. This design matches the
conventions from our literature review.

\textbf{Loop-Preserving $k$-hop Filtering.} CFGs can easily reach tens of
thousands of nodes, which is both prohibitively time-consuming for many layout
algorithms as well as unnecessary for a users' tasks. As we observed in the
task analyses and designs for CFGExplorer and CcNav, as well as the diagrams
in the literature, focusing on local structure is often sufficient, or even
preferred. Users may want to locate a loop or function in the code to begin
their search.

We thus built filtering down to a structure-aware subgraph directly into the
specification. The filter builds a subgraph in three stages. The filter
is initially defined by a list of specific nodes. An optional second stage
expands the subgraph to include all loops containing those nodes. Finally, the
subgraph is expanded to include all nodes within $k$-hops, where $k$ is a
tunable parameter. Additional parameters allow for further fine-tuning on the
number of nodes retrieved. These are described in the \textit{CFGConf}
documentation~\cite{CFGConfDocs}.

We chose a list for specifying seed nodes as a simple way to specify by hand
or via analytics scripts. We expect large graphs and complex analysis will
require the latter along with tuning drawing parameters. The integration of
drawing and filtering here is by design to speed up this process. During our
design iterations, one collaborator indicated they liked the tight workflow
loop afforded by this choice.

CcNav's filtering capabilities also used seed nodes and a $k$-hop
neighborhood, but did not preserve loop structures. Thus, loops could be lost
in filtering. We observed our collaborators frequently asking whether a node
was part of a loop, leading us to this new filtering approach. Additionally,
the CcNav filtering was a separate module from the laout, while
\textit{CFGConf} integrates the filtering with the layout specification.

\textbf{Boundary nodes.} To provide context in filtered graphs, we
draw nodes just outside of the filter, showing connectivity of the subgraph to
the rest of the CFG. We call these just-outside entities \textit{boundary
nodes}. Similar to stubs in Dynasty~\cite{eisner2006dynasty} and auxiliary
nodes in TensorBoard Graphs~\cite{wongsuphasawat2018tfgraph}, we represent
boundary nodes as small discs. When there are multiple boundary nodes
connected to an included node, they are aggregated and depicted as stacked
discs as shown in Fig.~\ref{fig:teaser}.

To draw boundary nodes, we let the graph layout algorithm place them the same
way as a regular node and then as a post-processing step move them closer to
the attached node. Since these nodes are generally on the outskirts of the
graph, moving them closer typically does not introduce new edge crossings.

Our visual design and layout was iterated on with collaborators. We showed
them mockup layouts that forced all boundary nodes to the borders of the
subgraph bounding box, but they considered that design too cluttered. We also
showed them designs with dotted line boxes instead of discs and sequences
instead of stacks. They rejected both of these in favor of the discs, with the
rationale that they attract less attention.

\textbf{Function collapsing.} While CFG experts may want to preserve certain
structures, they might deem others uninteresting
or inappropriate for their level of analysis. However, these uninteresting
data items can compose a significant portion of the visualization and
therefore impact the layout and its performance. For example, commonly called
library and utility functions, such as \texttt{printf} and \texttt{puts},
result in high-degree nodes due to all the disparate calls to them. This can
clutter and warp the layout significantly. However, these functions are often
not relevant to the analysis. 

\textit{CFGConf} can render nodes associated with a particular function as a
single node, effectively \textit{collapsing} it into a single entity. To
further declutter the depiction and decrease the effect of these nodes on the
layout, we duplicate collapsed nodes where called, again inspired by
Tensorboard. The collapsed nodes are drawn as dashed boxes, as shown in
Fig.~\ref{fig:singlefile} (bottom) and later in case study
Fig.~\ref{fig:ltimescollapseexample}. This depiction decreases edge crossings
and reduces focus on less important nodes while retaining context.  Function
collapsing acts as another structure-aware, context-preserving filter on the
drawing. 

Users can specify the collapsed functions as a list of names or by graph
properties in the CFG (or both in tandem). Graph properties considered are
node degree and the size of the function with respect to the active subgraph.
The rationale is that utility functions often have high-degree nodes (e.g.,
the entry and return nodes) because they are called frequently. Users can set
the degree at which to collapse. Similar to the filtering options, parameters
to tune the size of the subgraph specified are also available. Our
collaborators noted the importance of heuristics because enumerating a
comprehensive list of function names would be difficult.

We chose the name \textit{collapsing} because our collaborators found it the
most intuitive. We had originally called it ``splitting'' for the duplication
of nodes which also occurs, but this was considered a secondary step in the
layout alteration.

\subsection{System Design}

We built \textit{CFGConf} in Javascript for ease of integration into web-based
multi-view systems. \textit{CFGConf} parses the specification
using graphlib-dot~\cite{graphlibdot:online}. It uses GraphViz \textit{dot}
algorithm as the base layout engine for the extensive support and
customizations options. 
%(This is different from CFGExplorer and CCNav's layouts which are
%based on \texttt{dagre.js}~\cite{dagre}.) 
From the parsed specification,
we apply the customizations, using graphlib~\cite{graphlib:online} for
processing. As GraphViz was used a base, we can then output both \texttt{dot}
format and SVG. Element styling beyond \texttt{dot} specifications are done
using \texttt{D3.js}~\cite{bostock-d3-2011}.

The library includes a sample webpage for housing SVG examples. The page
supports panning and zooming the CFG. These demonstration interactions are
loosely-coupled so that \textit{CFGConf} can be more easily integrated into other
systems that support those interactions differently. \textit{CFGConf} is
available open-source at: \url{https://github.com/hdc-arizona/CFGConf}.

\begin{figure*}[htb]
\centering
     \begin{subfigure}[t]{0.21\textwidth}
         \centering
         \includegraphics[width=0.98\textwidth]{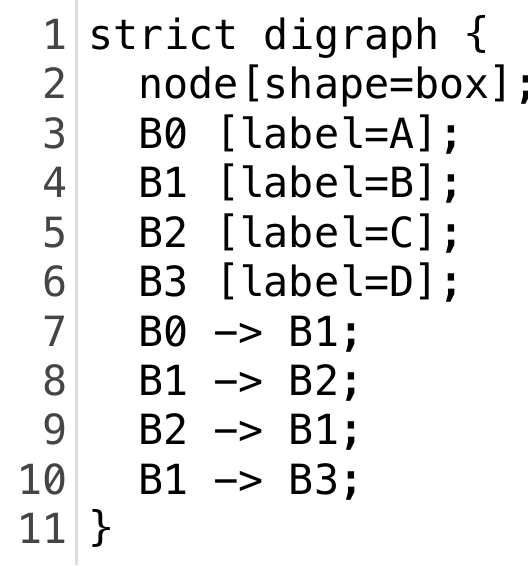}
	 \caption{\texttt{dot} specification}
         \label{fig:dotnoloop}
     \end{subfigure}
     \hfill
     \begin{subfigure}[t]{0.32\textwidth}
         \centering
         \includegraphics[width=0.98\textwidth]{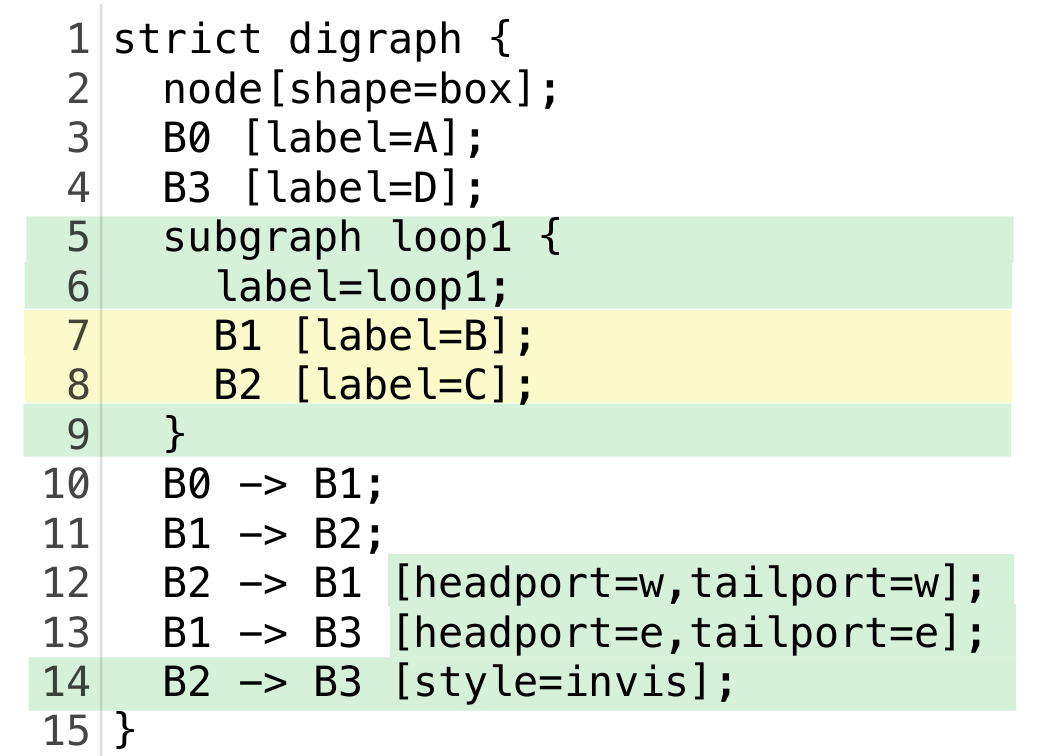}
	 \caption{\texttt{dot} specification with loops}
         \label{fig:dotloop}
     \end{subfigure}
     \hfill
     \begin{subfigure}[t]{0.24\textwidth}
         \centering
         \includegraphics[width=0.98\textwidth]{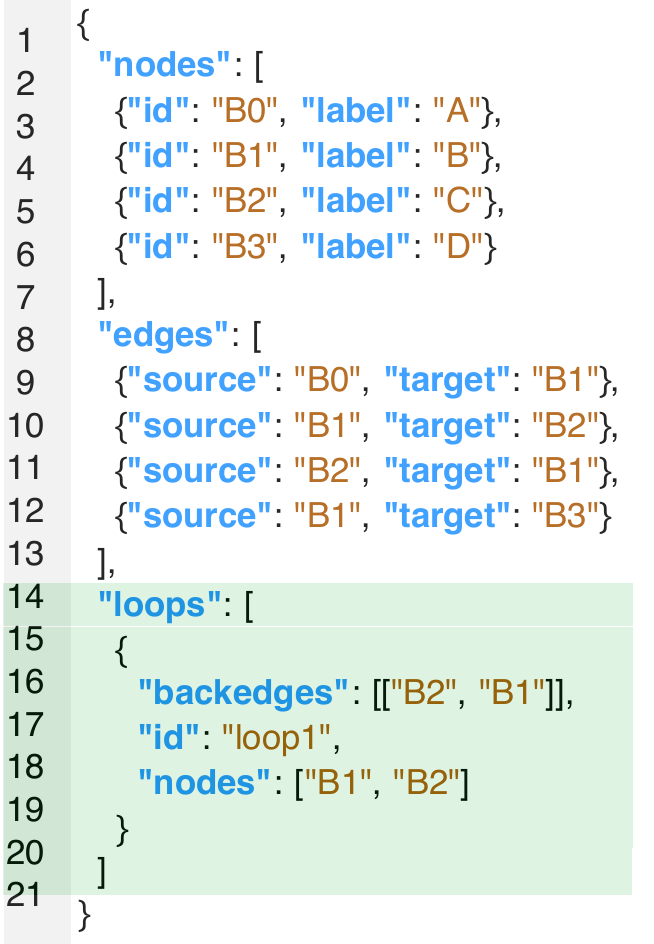}
	 \caption{\textit{CFGConf} specification}
         \label{fig:cfgconfloop}
     \end{subfigure}
     \hfill
     \begin{subfigure}[t]{0.20\textwidth}
         \centering
         \includegraphics[width=0.98\textwidth]{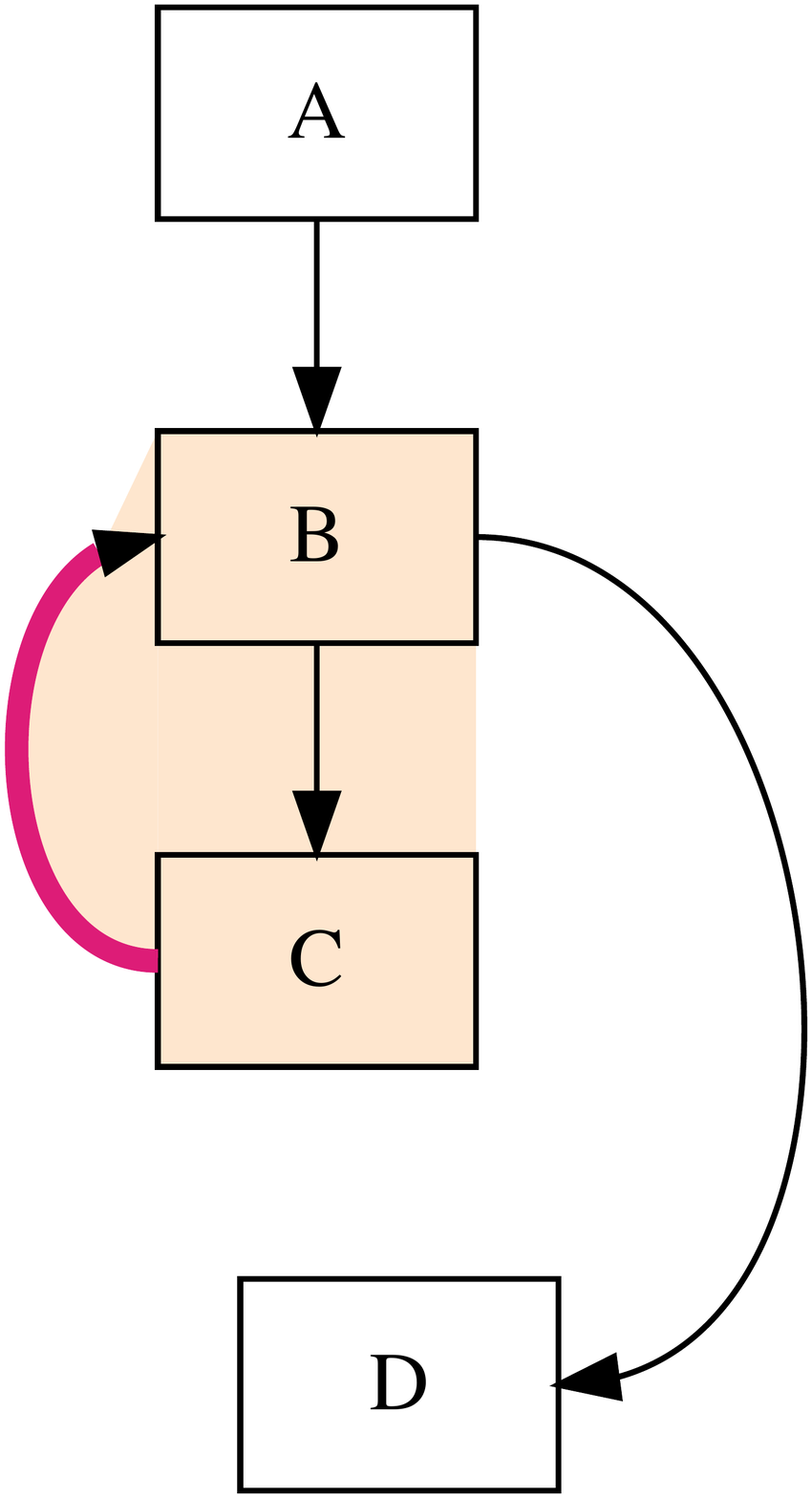}
	 \caption{drawing with loops}
         \label{fig:resultingloop}
     \end{subfigure}

\caption{Domain-specific loop visualization in \textit{CFGConf}'s JSON and
  \texttt{dot}. Lines specifying the loop are in green. Lines that require
  re-arranging from the non-loop case are in yellow. The \textit{CFGConf}
  specification uses domain language (e.g., \texttt{loops}) while 
  \texttt{dot} is lower level and requires creating a subgraph, rearranging nodes, adding
  invisible edges, and adding port routing.} 

\label{fig:cfgconfvsdot}
\end{figure*}

\subsection{CFGConf Specification Language Design}
\label{subsec:language_design}

\textit{CFGConf} is designed for computing systems experts to use 
stand-alone or via integration into multi-view visualization systems. To be
usable by this audience, it should be efficient, learnable, and expressive.
There are three main strategies in achieving these goals. First,
\textit{CFGConf} is based on widely-known, accessible formats JSON and
\texttt{dot}, so it leverages existing familiarity as well as the simplicity
of JSON. Second, it allows users to specify layout and rendering options in
terms of higher level program structures like functions and loops that are
directly related to their tasks. This higher level of abstraction further
keeps the specification concise. Third, \textit{CFGConf} requires only
topology and program structure data to run. Other parameters are
optional. If collapsing or filtering are specified without parameters, default values
are used, further decreasing the burden on the user.

During development, we used the cognitive dimensions of notation
(CDN)~\cite{cogdimnotation1989} framework as a check on the initial prototype
language specification. CDN are a set of design principles for programming
languages that have been used to evaluate other visualization toolkits, such
as ProtoVis~\cite{protovis2009}.  Two authors independently evaluated the
early \textit{CFGConf} design under CDN, verifying it rated ``Good'' or better
under the vast majority of heuristics before presenting it to collaborators.
The heuristics with lower scores were due to the limited scope of
\textit{CFGConf} and the trade-offs in leveraging JSON.

\textbf{Specifying CFG Data.} Node, link, and structure data are specified by
a top level keyword, \texttt{data}. \textit{CFGConf} supports specifying nodes and
links either through JSON or \texttt{dot} format, which is widely used in
computing contexts~\cite{isaacs2019graphterm}. Loop and function data is
specified through JSON only. Either format can be linked from a separate file
which leaves the tunable \textit{CFGConf}-specific parameters in a concise file.
%The JSON data may also be specified inline. 

While only structure information is required (e.g., \texttt{id},
edge \texttt{source} and \texttt{target}, node membership in
loops), style directives for specific nodes are optional, either using
parameters recognized by the \texttt{dot} format or by CSS classes.
Alternatively, global style rules may be set (see next section).

Fig.~\ref{fig:cfgconfvsdot} shows the CFG of a \texttt{while} loop and the
modifications required to draw it in a domain-specific fashion using the
\textit{CFGConf} JSON specification compared to \texttt{dot}. Adding loop
semantics to \textit{CFGConf} only requires adding the \texttt{loops} key while
\texttt{dot} specification requires creating a subgraph, moving the nodes,
adding invisible edges, and adding port routing. 

\textbf{Specifying Style and Layout Parameters.}
\textit{CFGConf} groups default global style directives and layout parameters
under a top level \texttt{rendering} keyword. Style parameters can be set for
\texttt{node}s, \texttt{link}s, \texttt{loop}s, and \texttt{function}s using
nested objects with these keywords. They are applied to all entities except
those overridden in the individual specifications under the \texttt{data}
keyword.  \textit{CFGConf} supports the style keywords (e.g., \texttt{shape},
\texttt{color}, etc.) of \texttt{dot} as well as setting a \texttt{class}
parameter for further styling with CSS.

The \texttt{function} specification determines whether the layout will group
nodes belonging to the same function in an enclosing rectangular boundary. For
\texttt{loops}, users have the option of not setting loop background colors.

\textbf{Specifying Selected Nodes and Filtering.}
Instead of rendering the full CFG, users can specify a subset of the nodes to
be shown. They may list all nodes directly or use a higher level of
abstraction by taking topological distance and program structures into
account. This feature is parameterized in a top level \texttt{filtering}
keyword as this feature determines {\em what} is shown rather than {\em how}
it is shown. Users specify a subset of nodes (\texttt{selectedNodes}). They
can then grow that subset using associated  loop data and graph distance or
limit the size of the resulting subgraph in terms of nodes.

\textbf{Specifying Function Collapsing and Duplication.}
\textit{CFGConf} allows users to specify heuristics for minimizing the
saliency of nodes associated with uninteresting functions while retaining
context. The \texttt{collapsingRules} keyword is nested under
\texttt{function} in the \texttt{rendering} keyword, as this feature affects
layout and presentation. 

The collapsing heuristics include the in- and out-degree of the function. The
rationale is that utility functions are called from many places, and thus will
have at least one high degree node. The collapsed set can be further adjusted
by setting limits on the number of nodes in a collapsed function. 
Functions containing loops are not collapsed due to their interest as
waypoints in the analysis. Users retain the ability to override these
heuristics for specific functions by placing them in the
\texttt{alwaysCollapseList} or \texttt{neverCollapseList}.

\section{Examples and Integration Case Study}
\label{sec:examples}

We illustrate the expressiveness of \textit{CFGConf} for specifying CFG
drawings in the following examples (Sect.~\ref{sec:rawexamples}). Full data
files are included in the supplementary materials. We also discuss the
integration of \textit{CFGConf} into CcNav, replacing the previous layout, to
demonstrate its usability as a modular component in a multi-view system
(Sect.~\ref{sec:ccnavexample}).

\subsection{Stand-Alone Examples}
\label{sec:rawexamples}

We discuss two stand-alone CFG examples taken from real programs of interest
to our collaborators. These examples demonstrate how domain-specific drawings
can be expressed in \textit{CFGConf}'s JSON specification.

\begin{figure}[!htb]
\centering
\includegraphics[width=0.3\textwidth]{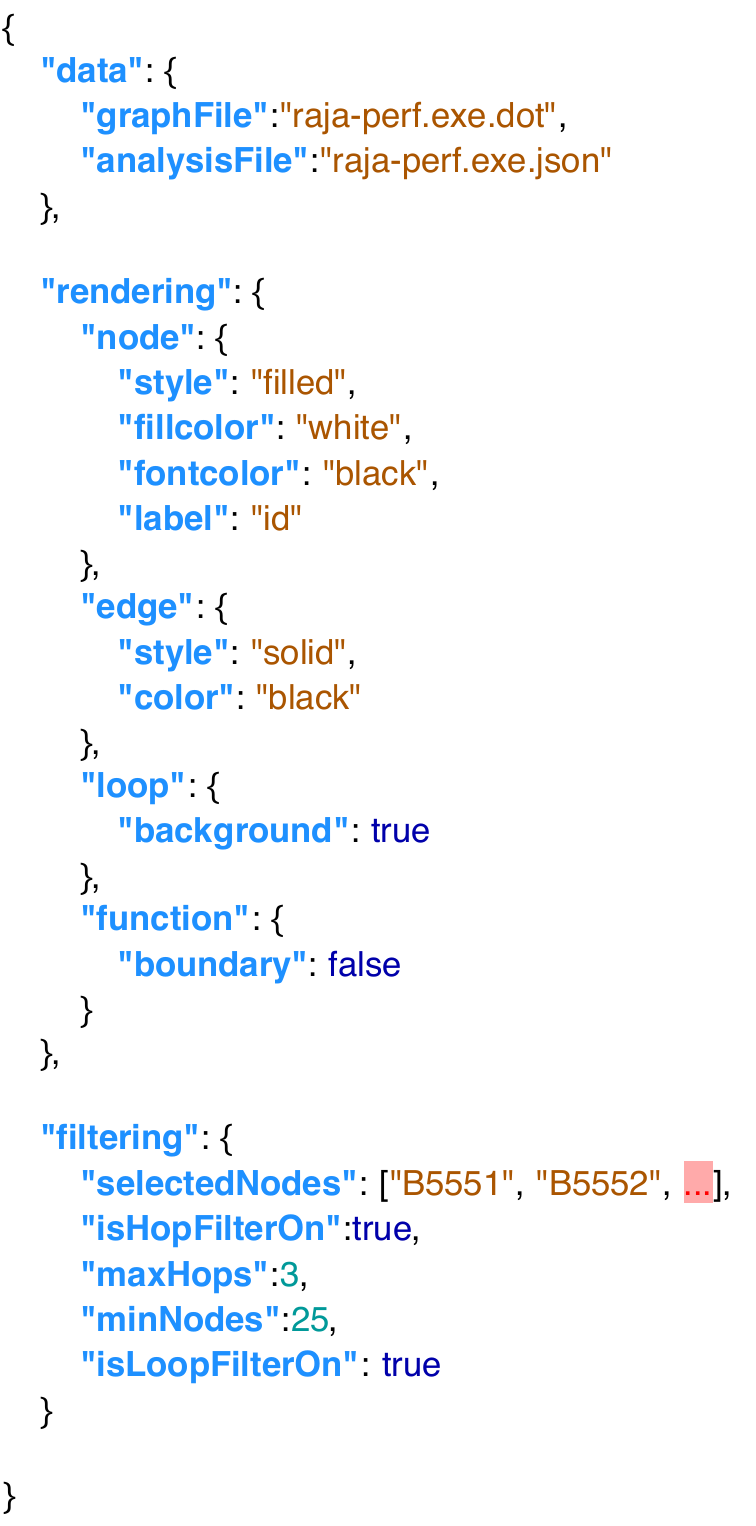}
\hfill
\includegraphics[width=0.14\textwidth]{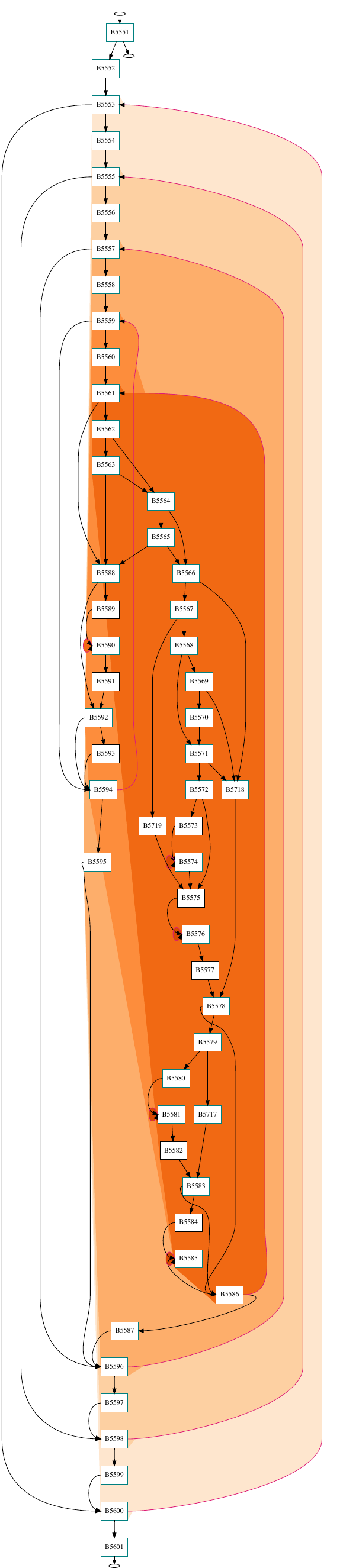}
  \caption{The subgraph of the LTIMES CFG selected by \textit{CFGConf}'s filtering specification. Though the full graph has 18.7K nodes, we only visualize a subgraph based on nodes provided to \texttt{selectedNodes}.} 
\label{fig:rajaperfltimes}
\end{figure}

\textbf{Filtering to Loops of Interest.} This example shows how
filtering is used to visualize structures of interest in the code. The data
was collected from the LTIMES application of RAJAPerf~\cite{rajaperf} using
\texttt{optparser}~\cite{devkota2021ccnav}. The collected CFG has over 18,700
nodes, making it infeasible to lay out in full with a Sugiyama-style algorithm.
However, our collaborators want to examine a particular nested loop, so we use
filtering to focus on that loop. Fig.~\ref{fig:rajaperfltimes} shows the
\textit{CFGConf} specification and resulting visualization.

The \texttt{filtering} feature is enabled with \texttt{isHopFilterOn} and
seeded with node \texttt{id}s of interest in \texttt{selectedNodes}, which are
elided here for space. The drawing shows the selected nodes with a teal
border. These nodes are related to a line of code in the loop. They were
input manually for this example, but a multi-view interactive system like
CcNav supports selecting these nodes through linked brushing of a source code
view. 

The \texttt{isLoopFilterOn} key ensures any loops containing the
\texttt{selectedNodes} are included in the subgraph. This feature is important
in this example as our collaborators focus on loops for optimization. The
subgraph is further expanded to all nodes within three hops, as set by the
\texttt{maxHops} key, while \texttt{minNodes} ensures at least 25 nodes are
returned should the \texttt{maxHops} selection be too small. Nodes just
outside the subgraph are shown as small discs. As these are only seen at
the top and bottom of the image, we infer there are no calls to functions
outside the loop.

The specification also includes global styling options for \texttt{node},
\texttt{edge}, \texttt{loop}, and \texttt{function} entities under the
\texttt{rendering} keyword. Here the \texttt{node} and \texttt{edge} styles
follow the \texttt{dot} language, with the addition of a
\texttt{label} parameter in \texttt{node} which directs \textit{CFGConf} to
use the node's \texttt{id} as a label. Additionally, \texttt{loop} is set to show extent
background colors, and \texttt{function} boundaries are not drawn.

\begin{figure}
\centering
  \includegraphics[width=0.99\columnwidth]{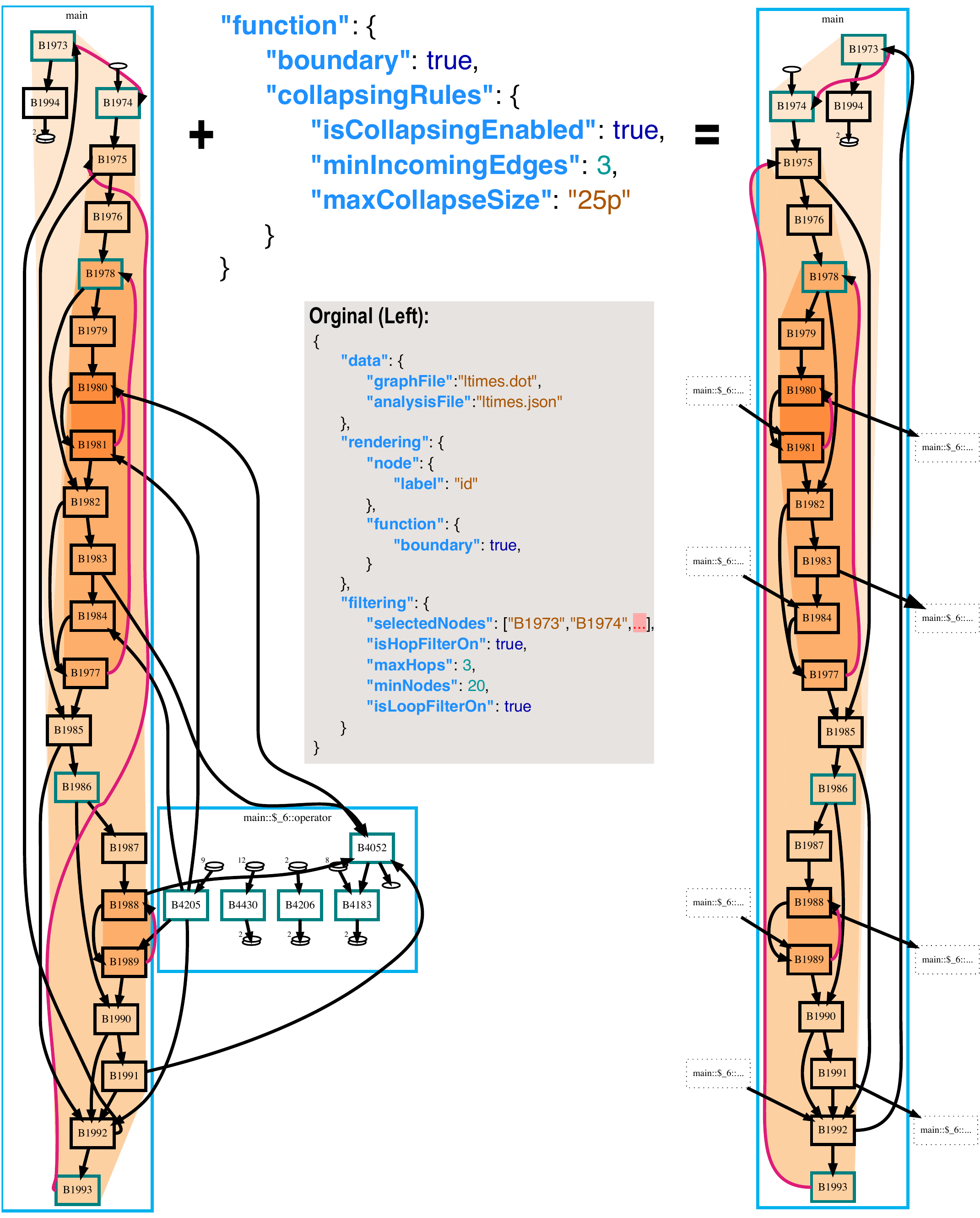}

  \caption{Drawing of a filtered graph with and without function
  collapsing enabled. 
  Left: As the function \texttt{main::\$\_6::operator} is called
  multiple times from the loop, there are several incoming (function call) and
  outgoing (call return) edges. The small stacked discs indicate several nodes from
  \texttt{main::\$\_6::operator} were not included in the filter. Right:
  \texttt{main::\$\_6::operator} has been collapsed and replicated as dotted
  rectangles, de-emphasizing it but showing where it connects.}
  \label{fig:ltimescollapseexample}

\end{figure}

\textbf{Collapsing and Duplicating functions.} This example demonstrates
simplifying the graph through the function collapsing feature. As in the
previous example, we use the LTIMES application, but filtered with a different
set of \texttt{selectedNodes}. Fig.~\ref{fig:ltimescollapseexample} shows the
specification and resulting drawings with and without the collapsing rules.

Collapsing is set using the \texttt{collapsingRules} object.
\mbox{\texttt{minIncomingEdges}} is set to three, indicating that functions
with nodes with three incoming edges (i.e., called from at least three
places) should be collapsed. \texttt{maxCollapseSize} is set to
\texttt{25p}, which limits collapsing to functions that comprise less than
25\% of the filtered graph. This preserves functions
representing a significant portion of the graph.
 
The collapsed version draws duplicated rectangles with a dotted border for
\texttt{main::\$\_6::operator}. Without collapsing on, several nodes internal
to the operator are drawn and grouped in a function boundary. As this operator
is called in several places, there are multiple incoming edges to B4052 and
multiple return edges from B4205. The multiple stacks of discs indicate the
operator contains several nodes not pulled in by the filter.

\subsection{Integration with CcNav}
\label{sec:ccnavexample}

CcNav~\cite{devkota2021ccnav} is a multi-view visualization system designed to
support program analysts in understanding what choices the compiler made. To
assist this task, CcNav includes views of the source code, disassembled binary
code, the CFG, a hierarchy of loops, a graph of function calls, and a
hierarchy of inlined functions. These six views have linked highlighting and
subselection, conditioned on the instructions involved. 

In the CFG, users may select nodes by clicking or with a multi-node lasso.
The other views will redraw based on the assembly instructions associated with
the selected nodes. Selections in other views will cause the CFG view to draw
a filtered subgraph of nodes with the selected instructions. 

We replaced the previous custom CFG component in CcNav with \textit{CFGConf},
maintaining these linking features. The updated system has been released
publicly on Github~\cite{ccnavgithub}. This integration adds the
loop-preserving filtering of \textit{CFGConf} over CcNav's original
$k$-hop-only filtering.

The integration was performed primarily (95\% effort) by a software engineer
who had not been involved in the implementation of \textit{CFGConf} and had
not originally implemented the linking portion in CcNav. He had previously
managed CcNav's file selection features, windowing management features, and
deployment on our collaborator's internal network.

The integration took approximately five days of development time, spread over
several weeks of real time, of which the vast majority (approximately 95\%)
was spent determining how both CcNav and \textit{CFGConf} work and which events
should trigger which actions. These activities were split evenly between CcNav
and \textit{CFGConf}. The software engineer further noted, ``There was very
little coding involved.'' This experience demonstrates that \textit{CFGConf}
can be used in multi-view visualization systems with minimal code and
reasonable development effort.

\section{Evaluations}
\label{sec:evaluation}

In addition to the examples in the previous section, we further
evaluate \textit{CFGConf} with a reproduction
study~\cite{reflectionVisEval2018} in which seven participants are asked to
use \textit{CFGConf} to produce specified drawings (Sect.~\ref{sec:userstudy})
and a heuristic evaluation using the cognitive dimensions of notation
(Sect.~\ref{sec:cogdimeval}).

We chose to evaluate with a combination of a participant study, illustrative
examples, example of integration, and heuristic evaluation as each method has
evaluatory strengths and limitations. The illustrative examples exist to
demonstrate expressiveness. The participant study evaluates the usability of
the specification and the utility of both the visual design and choice of
overall library design---informing to what extent potential users can
replicate CFG drawings under various scenarios should they find the library on
their own  as well as suggesting which if any design choices are helpful to
them and their future tasks.

The integration study exercises the goal of making it easier to include
domain-specific CFG drawings in a more complex visualization system, so others
can create systems like CFGExplorer and CcNav more geared to their larger
analysis goals. Finally, the heuristic evaluation adds another perspective on
usability and effectiveness of the specification which may be more systematic
and comprehensive in evaluating the notation but lacks the integrated and
human outcomes that the participant study seeks out.

\subsection{Reproduction Participant Study}
\label{sec:userstudy}

We summarize the reproduction study methodology and results. The study
materials and raw survey results are available in the supplemental materials
(\url{https://osf.io/vm3pc}).

\subsubsection{Methodology} 

The study was conducted asynchronously: participants were given instructions,
asked to perform tasks, and then submit the \textit{CFGConf} specifications
they created and complete a survey. They were permitted to do the tasks in one
or many sessions. The study was designed to take no longer than 90 minutes
based on piloting with two participants. Participants were provided a link to
study instructions, the \textit{CFGConf} repository, and the \textit{CFGConf}
wiki, which includes full documentation and a quickstart guide. We chose an
asynchronous approach with no training (other than the documentation available
to them) to better simulate someone using \textit{CFGConf} ``in the wild'' and
to provide flexibility to encourage more participation.

\textbf{Tasks and Survey.} To keep the study length reasonable, we limited the
study to four tasks and an eight-question survey. We summarize the tasks below.

\begin{enumerate}[font={\bfseries},label={T\arabic*.}]
\itemsep=0.25ex
\item Reproduce a given image of a four-node CFG with a \texttt{while}
loop.
\item Use both \textit{CFGConf} and \textit{dot} to draw the given CFG and comment
on the resulting drawings.
\item Filter and draw the CFG according to given text criteria.
\item Draw the CFG collapsing functions matching text criteria.
\end{enumerate}

Graph and structure files were given for tasks T2-T4 so the participants only
had to write the drawing specification. In T1, they had to write the graph
specification as well.  The survey asked for a summary of strategies used and
then general feedback regarding \textit{CFGConf}.  The task prompts and survey
questions are given in full in the supplemental materials
(\url{https://osf.io/vm3pc}).

\textbf{Participants.} We recruited seven participants (P1-P7) through
contacts from prior projects and solicitations to instructors of compilers
courses. P1-P4 had previous experience with CFGs and had visualized them for
their analysis. P5 had graph drawing experience only. P6 and P7 were graduate
students who had recently taken a course in compilers, but do research in
other computing areas. None were involved in the design of \textit{CFGConf},
but two (P2, P3) were domain collaborators in the CFGExplorer design.

\subsubsection{Results and Survey Feedback}

P1, P3, P4, and P6 completed all tasks. P2, P5, and P7 completed the first
two, though P7's was missing a parameter. For the final two tasks, P2 did not
realize there was an \texttt{analysisFile} option and thus could not load the
loop and function data. P5 and P7 got stuck trying to visualize the large
graph prior to filtering.  

\textbf{Feedback on \textit{CFGConf} drawing and features.} In the comparison
task, T2, participants noted that \textit{CFGConf} made it easier to see
structures (P1, P4, P5, P6). P1 wrote they ``found structure to be immediately
clearer with CfgConf.'' P6 wrote, ``The graph generated by CFGConf is way more
organized and easy to understand than the raw pdf graph'' and noted the back
edge highlighting. P5 wrote, ``With highlighting inner loops, etc., there's at
least immediate context where I think I could start to figure out what is
actually going on. With the raw DOT output, I expect there'd be a long
explicit step where I'd be manually trying to figure out the loop hierarchy.''

In further comparison with \textit{dot}, P1 and P3 both mentioned they
preferred the way LLVM's \textit{dot} file handles true/false edges for conditionals.
P1 referred to this as a ``minor quibble.''

Participants also liked the filtering (P3, P4), collapsing (P3), and rendering
(P4) options, as well as the overall layout and structure highlighting (P1) in
\textit{CFGConf}. P3 explained that the filter works similar to how they would
have done it manually and that they have often run into issues rendering large
graphs. They wrote ``For me, I think this is perhaps the most useful feature
in cfgConf.'' They also wrote effusively regarding the collapsing feature,
noting ``I've done some work where this would be super useful to collapse
commonly used functions that I didn't care about much.''

\textbf{Usability of \textit{CFGConf}.}
Regarding what they found easy about using \textit{CFGConf}, participants
picked out the setup (P1, P4), the producing drawings (P3, P4,
P6, P7), and the ability to use their own files (P2, P4, P5). P4 wrote it was
an ``intuitive interface for describing, and loading graphs and also Setting
filtering and rendering options.''
P5 liked how \textit{CFGConf} could use an existing graph file or
used on its own, allowing the creation of diagrams without data.

Regarding difficulties, the most prevalent comment was the lack of error
handling or error messages (P1, P4, P6, P7). Both P1 and P6 noted in the large
graph case, it was unclear if there was an error or the algorithm was just
taking a long time. 

Participants also noted the zoom behavior in the web application (P1) and the
ease of making syntax errors in JSON (P2), the verbosity of JSON (P2), and the
large amount of documentation (P5) as difficulties. P5 explained the
documentation did not match their strategies of skimming or Control-F search,
though noted their lack of familiarity with the domain could be a contributor.

\textbf{Suggested improvements.}
There were multiple calls for better error handling, error detection, and
error messages (P4, P6, P7). P7 further suggested integration with an IDE. 

P5 suggested adding more background and design
rationale to the documentation to aid Control-F search. P1 and P4 expressed
confusion with how some filtering parameters interact, which may also
suggest the documentation can be improved.

Other suggestions for included improved zooming (P1), error messages (P1),
minor rendering improvements (P1), and more options in function collapsing
(P3). P3 and P4 suggested adding more dynamic features such as interactive
filtering and path tracing. Our intent with \textit{CFGConf} was that such
features could be built by systems using \textit{CFGConf}.

\textbf{Scenarios of Use for \textit{CFGConf}.}
When asked about scenarios in which they would use \textit{CFGConf}, five
participants (P1, P3, P5, P6, P7) described scenarios. P2 and P4 said they had
no scenarios ``currently'' and ``at this time'' respectively.

P3 said they currently work with CFGs and would use it to validate and debug
their work.  P1 wrote they would ``definitely use CFGConf'' in situations with
large \texttt{dot} files or when they wanted to see structure.

P6 said they would have used \textit{CFGConf} had they known about it
while they were implementing a compiler for a course. They further explained
that they ``got some insights regarding visual aesthetics by using
[CFGConf]'' which they might apply to a network visualization in a different
domain.

P7 wrote \textit{CFGConf} would be useful for analyzing code structure for
optimization or designing a modular library to check dependencies between
functions.  P5 noted they do not work in program flow, but would use it in
situations where they expected nested loop structures or similar topology.

\subsubsection{Discussion} 

All of the participants were successful in creating graph drawings in the
first two tasks, but three were unsuccessful in the second two tasks. Despite
these difficulties, participants generally found \textit{CFGConf} easy to use.
The reasons for the task failures---attempting to draw too large a graph
without filtering and misunderstanding the type of file---further indicate the
need for error detection and error messages, which participants also directly
expressed in the follow up survey. The graph drawing successes and ease of
use feedback despite errors is validating of the language design, with the
exception of error-related aspects.

The biggest exception regarding ease of use was P5, who despite extensive
graph drawing experience did not have CFG experience. This suggests
\textit{CFGConf} has a strong domain-mapping, which seems to help domain
experts, but creates a trade off with non-expert users. 

The participants with domain knowledge liked how the \textit{CFGConf} visual
design preserved structures and had options for de-cluttering, such as
collapsing. Filtering was brought up by multiple participants as a much
desired feature. Most suggestions outside of error handling were minor
improvements or building on the UI, rather than the language or the visual
design. This feedback serves to validate the \textit{CFGConf} features and
visual design.

\subsubsection{Limitations and Threats to Validity.} We sought participants
familiar with both CFGs and drawing who could volunteer 90 minutes, thus
limiting the number we were able to recruit, and in turn limiting the
generalizability of the feedback. Furthermore, five of the participants
(P1-P5) each know one or more authors, which may have biased their feedback.

Only four of the seven participants created filtered and collapsed graphs,
further limiting our findings regarding those features. The points of failure
for the three who did not complete the final two tasks suggest limitations in
the asynchronous study design. Though contact information was available and
the participants were told to contact the research team if they got stuck,
only one user did so. P5 wondered if the study was testing if the users could
figure out the documentation, which was not the intent of the study design. 

We chose a reproduction study to evaluate the usability of the
\textit{CFGConf} specification language. It tested how people are able to
match given criteria, but while it asked for speculative feedback regarding
future projects, it did not test how people might use \textit{CFGConf} for
their own problems. Long-term deployment feedback is needed to assess
\textit{CFGConf}'s applicability and understand more generalizably the
strengths and weaknesses of the language.

\subsection{Cognitive Dimensions of Notation}
\label{sec:cogdimeval}

We revisited the CDN heuristics (Sect.~\ref{subsec:language_design}) using the
most recent iteration of \textit{CFGConf}. We summarize our findings here. A
discussion of \textit{CFGConf} under each heuristic is available in the
supplemental materials (\url{https://osf.io/vm3pc}).

We assigned a Very Good or Excellent rating to seven of the thirteen CDN
heuristics considered, indicating a good match. The language is {\em
Consistent} in its own design and with that of other standards as it draws
from \texttt{dot} and CSS. The specification also matches with the problem
language, as noted by study participants ({\em Closeness of Mapping}). 

By virtue of its JSON syntax, \textit{CFGConf} can be written in any order
(avoiding {\em Premature Commitment}) and once the data is present, can be run
without more settings as long as the graph is not too large ({\em Progress
Evaluation}). Changing a key-value pair other than IDs requires few other
changes ({\em Viscosity}).  The data, rendering, and filtering features are
separated, supporting {\em Role-Expressiveness}. This separation, along with
the single-file nature, helps users locate pieces of the specification and
compare them ({\em Visibility}). 

We made a trade-off with some of the heuristics to gain the familiarity and
support that comes with JSON. Specifically, JSON does not have direct
commenting support, limiting its fulfillment of the {\em Secondary Notation}
heuristic. It can also be {\em Error-Prone} in terms of missing commas and
quotes, as one study participant mentioned. Using JSON to specify the graph
data can also be more verbose than \texttt{dot} ({\em Diffuseness}). 

Graph operations are another source of cognitive difficulty. Layout algorithms
are complex enough that users cannot perfectly match specification changes
with layout changes ({\em Hidden Dependencies}) and operations like filtering
may require careful thought ({\em Hardness of Mental Operations}).
\textit{CFGConf} also does not support the creation of one's own structures,
e.g., conditionals as noted by study participant P3, and thus does not support
{\em Abstraction}.

\section{Reflections}
\label{sec:discussion}

We reflect upon the \textit{CFGConf} project and insights gained regarding the
combination of filtering and drawing and the results of the multi-modal
evaluation.

\subsection{Combining Filtering and Drawing} Feedback from participants
regarding filtering was enthusiastic. We hypothesize that combining filtering
with drawing fits debug workflows that typically use \texttt{dot}. This
follows from a use case discussed by one of our domain expert collaborators,
an author on this work, who suggested times where they might want to tweak the
graph quickly using a text editor rather than through an interactive
visualization system.

We suspect the need to filter and draw in a lightweight scripting style may
extend to other domains, but the right level of abstraction and trade offs
between flexibility and simplicity is yet unknown. For example,
\textit{CFGConf}'s filtering features are limited, but have domain-aware
aspects. Comparing \textit{CFGConf}'s filtering to a workflow using a general
graph query language, like Cypher~\cite{cypher:online}, may help guide design
needs for domain-specific solutions.

\subsection{Multi-modal Evaluation} Visualization libraries and specifications
are challenging to evaluate~\cite{pu2021evaluation}. We took a multi-modal
approach with (1) examples to demonstrate expressiveness, (2) an integration
case study to validate \textit{CFGConf}'s use in multi-view visualizations,
(3) a participant study to verify usability and usefulness to the target audience,
and (4) heuristic evaluation using the cognitive dimensions of notation to
assess the specification language in terms of widely accepted design
principles. Our rationale is that each of these strategies evaluates
different, though related, goals of \textit{CFGConf} and to some extent covers
the limitations of the others. For example, while neither the CcNav case study
nor the user study complete captures how \textit{CFGConf} may be used ``in the
wild,'' they represent different real-world scenarios, CFG drawing in a large
visualization project (CcNav) and lightweight CFG drawing by users unfamiliar
with the system (user study).

The results of these evaluations complemented each other, further
strengthening our findings. The successes and struggles of participants in the
user study were consistent with our cognitive dimensions assessment. The level
of mastery the participants attained in 90 minutes is consistent with the
multi-day effort required for the CcNav integration.

\section{Conclusion}
\label{sec:conclusion}

We presented a domain-specific set of visual designs for drawing control flow
graphs and a language for specifying them, implemented in the library
\textit{CFGConf}. Our approach focuses on preserving and prioritizing
domain-specific structures such as loops and functions and common tasks such
as identifying those structures, filtering for them, and wrangling
(collapsing) them. This approach was grounded in a review of drawing styles in
domain literature, prior visualization projects for CFGs, and an iterative
design process with domain experts.  

Through four separate types of evaluation, we demonstrated that the visual
design and language of \textit{CFGConf} enables users to produce intuitive
drawings of CFGs that support common tasks of identifying structures, focusing
on what's interesting while retaining context. Through a case study with
CcNav, we showed \textit{CFGConf} is viable for integration into multi-view
visualization systems. Feedback from domain experts in our study suggests this
higher level of abstraction matched the logic of our target users, improving
the ease-of-use and the resulting drawings.

\section*{Acknowledgment}
We thank our study participants for their valuable time and the LLNL LEARN
project, LLNS B639881 \& B630670, and NSF IIS-1844573 for supporting this
research. This work performed under the auspices of the U.S. Department of
Energy by Lawrence Livermore National Laboratory under Contract
DE-AC52-07NA27344. LLNL-CONF-838042. 

\bibliographystyle{IEEEtran}  
\bibliography{vissoft}

\end{document}